\documentclass[fleqn,twoside]{article}
\usepackage{espcrc2}
\usepackage{epsfig}
\usepackage{bm}

\usepackage{graphicx}
\usepackage[figuresright]{rotating}

\newcommand{\AmS}{{\protect\the\textfont2
  A\kern-.1667em\lower.5ex\hbox{M}\kern-.125emS}}

\title{The Casimir Energy Paradox of the QCD String}

\author{K. Jimmy Juge\address{Institute for Theoretical Physics, University 
   of Bern, Sidlerstrasse 5, CH-3012 Bern, Switzerland},
Julius Kuti\address{University of California at San Diego,
La Jolla, CA 92093, USA}
\thanks{Talk presented by J. Kuti.}
and	
Colin Morningstar\address{Carnegie Mellon University,
Pittsburgh, PA 15213, USA}}

\begin{document}
\begin{abstract}
It is widely thought that the 
early onset of the asymptotic Casimir energy with unit
conformal charge signals bosonic string formation of the confining
flux connecting a static quark-antiquark pair in QCD. 
This is observed on a scale
where most of the string eigenmodes do not exist and the few
stable modes above the ground state are displaced.
Hints for the resolution of this paradox are suggested.
\vspace{1pc}
\end{abstract}

\maketitle

\section{Puzzle in the QCD String Spectrum}

A new analysis of the fine structure in the QCD string spectrum was presented
at Lattice 2002~\cite{JKM}. Shortly afterwards, two papers were
submitted for publication with focus on complementary aspects
of the same problem~\cite{JKM,LW1}. 

\begin{figure}[t]
\epsfxsize=2.8in
\epsfysize=4.4in
\epsfbox{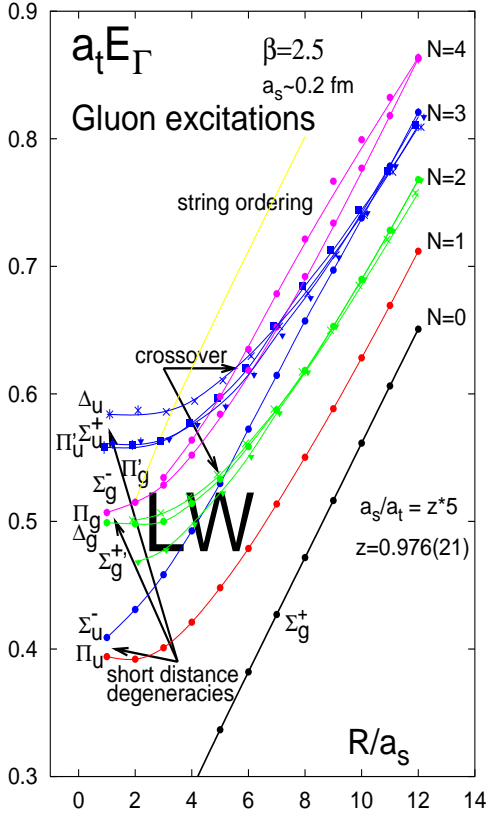}
\vskip -0.3in
\caption{Short distance degeneracies and crossover
in the QCD string spectrum from Ref.~\cite{JKM} where the notation
is explained. The symbol LW indicates the ${\rm R}$ range of ${\rm C_{eff}(R)}$
in Ref.~\cite{LW1}.
}
\label{fig:fig1}
\vskip -0.3in
\end{figure}

Ref.~\cite{JKM} reported a comprehensive study
of the QCD string spectrum as the quark--antiquark separation ${\rm R}$
was varied in the range ${\rm 0.2~fm < R < 3.0~fm}$ (Fig.~\ref{fig:fig1}).
On the shortest length scale, the excitations were consistent 
with short
distance physics without string imprint in the spectrum.
A crossover region below 2 fm was identified 
with a dramatic rearrangement of the level orderings.
On the largest length scale of 3 fm, the spectrum exhibited
string-like excitations with asymptotic ${\rm \pi/R}$ string gaps
split by a fine structure. 
It is quite remarkable
that the torelon spectrum, which is free of end effects,
exhibits a similar fine structure, as reported for the first time at
this conference~\cite{JKMMP}. 

In a complementary study~\cite{LW1}, the Casimir energy 
and the related effective conformal charge,
${\rm C_{eff}(R) = -12R^3F'(R)/(\pi(D-2))}$,
were isolated where ${\rm F(R)}$ is the force
between the static color sources and D is the space-time 
dimension of the gauge theory. 
With  unparalleled accuracy, ${\rm C_{eff}(R)}$ was determined for the gauge
group SU(3) in three and four dimensions
in the range ${\rm 0.2~fm < R < 1.0~fm}$ below the
crossover region of the string spectrum. It was suggested that 
the rapid change of the effective conformal charge from what
is expected in the running Coulomb law to ${\rm C_{eff}(R) \approx 1}$ well 
below 1 fm is a signal for early bosonic string formation.
The results are surprising because
the scale $\rm R$ is not large compared with the expected width of the
confining flux, and, perhaps more quantitatively, the string imprint 
in the Casimir energy is
observed in the ${\rm R}$ range where the spectrum exhibits complex non-string
behavior, as shown in Fig.~\ref{fig:fig1}. 

After a long series of comprehensive studies,
we report here new results in the three dimensional Z(2)
gauge model and a simple resonance model for
a better understanding of the seemingly paradoxical situation.

\section {String Formation and $\mathbf{C_{eff}}$ in Z(2) Model}

\begin{figure}[h]
\hskip -0.3in
\epsfig{file=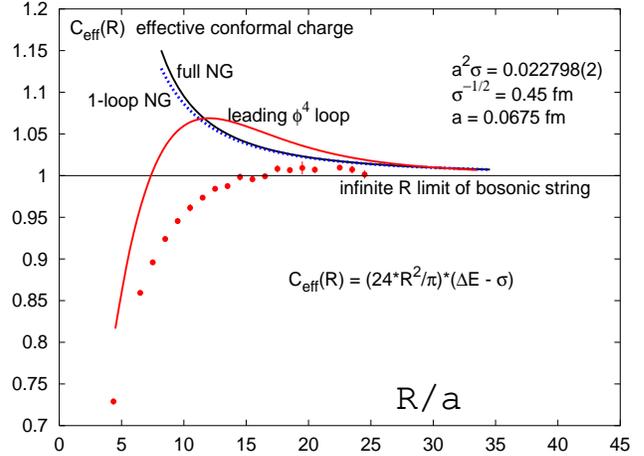,height=3.5in,width=2.5in,angle=-90}
\vskip -0.3in
\caption{The red points are from high precision Z(2)
simulations. The solid black curve with NG label is the full NG
prediction, ${\rm C_{eff}(R) = 1 }$ is the asymptotic string result (tree-level NG),
the dashed blue line is the 1-loop NG approximation. 
The solid red line represents the analytic
first term in the ${\hbar}$ expansion.
}
\label{fig:fig2}
\vskip -0.3in
\end{figure}

\begin{figure}[h]
\hskip -0.2in
\epsfig{file=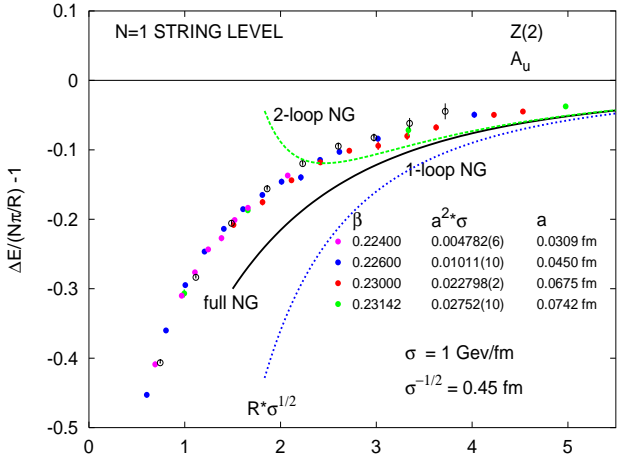,height=3.4in,width=2.5in,angle=-90}
\vskip -0.3in
\caption{The energy gap ${\rm \Delta E}$ above the ground state is plotted
as ${\rm \Delta E/(N\pi/R) - 1}$ to show percentage deviations   
from the asymptotic string level for ${\rm N=1}$.  Several Z(2) simulations
with cyan, blue, red, and green points
are combined with good scaling properties. The open circles 
represent D=3 SU(2) results after
readjusting the ratio of the string tension $\sigma$ to the glueball mass
in Z(2). The black line is the full NG
prediction, the dashed blue and green lines are 1--loop and two--loop
NG approximations, repectively.
}
\label{fig:fig3}
\vskip -0.3in
\end{figure}

References to earlier work on 
the three-dimensional Z(2) gauge model can be found
in a recent paper on the finite temperature
properties of the Z(2) string~\cite{CHP}.
The well-known dual transformation of the model to Ising
variables facilitates very efficient
simulations and analytic studies of the Casimir energy and the string
spectrum in the
${\rm \Phi^4 }$ field theory setting.
This is illustrated in Figs.~\ref{fig:fig2} and~\ref{fig:fig3}.
We applied the simple definition 
${\rm C_{eff}(R) = -24R^2(E'(R)-\sigma)/(\pi(D-2))}$
with the string tension $\sigma$ determined in high precision
separate runs from the ground state of long torelons.
The simulations in Fig.~\ref{fig:fig2} are compared with
the predictions of the Nambu-Goto (NG) string model
and the analytic first term in the $\hbar$ expansion of the 
equivalent ${\rm \Phi^4}$ field theory setting. 
The NG spectrum with fixed end boundary conditions 
in $D$ dimensions was first 
calculated in Ref.~\cite{arvis} with the result
${\rm  E_N = \sigma R ( 1 - \frac{D-2}{12\sigma R^2}\pi+ \
  \frac{2\pi N}{\sigma R^2} )^{\frac{1}{2}} }$ where N=0 is the string 
ground state.
Although there exists an inconsistency in the quantization of
angular momentum rotations around the ${\ q\bar q}$ axis at finite
${\rm R}$ values unless ${\rm D=26}$, the problem asymptotically 
disappears in the
${\rm R \rightarrow\infty}$ limit~\cite{arvis}.
It has been expected that predictions for strings emerging
from field theory and their effective bosonic string description
at large ${\rm R}$ will be similar to that of the NG model with corrections
which are increasingly important at smaller ${\rm R}$.

The field theory calculation of ${\rm C_{eff}}$ includes the classical 
energy of the string-like 
soliton being formed and the sum of zero-point energies
which are dominated by glueball scattering states in the bulk! The few
stable and displaced ``string-like" modes are also contributing.
How scattering states might conspire to produce
${\rm C_{eff}\approx 1}$ is further illustrated below in a simple
resonance model of massless scalar field theory.
Our results are closer to the leading order field theory calculation
and deviate substantially from
the predictions of the NG string model, particularly at smaller
${\rm R}$ values. This differs from the tantalizing findings of 
Ref.~\cite{CHP} where simulations
of the finite temperature Z(2) string were presented to be in
agreement with the 1--loop NG string model. 
For further illustration,
the first string excitation in Fig.~\ref{fig:fig3} is compared
with the NG model and its loop expansion which 
completely breaks down below 1 fm.
The leading order field theory calculation, which does not assume string
formation, accounts for the shape of the spectrum quite well.

\section{Simple Resonance Model}
\begin{figure}[h]
\hskip -0.3in
\epsfig{file=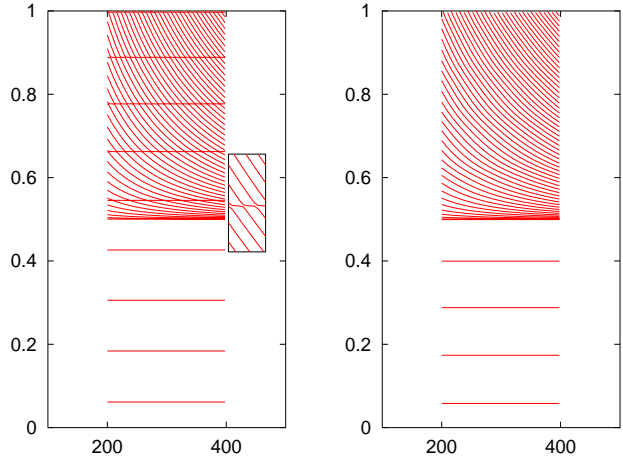,height=3.5in,width=2.5in,angle=-90}
\vskip -0.3in
\caption{ For ${\rm M=\frac{1}{2} }$, energy leveles ${\ E_n(L) }$ 
(vertical coordinate) 
are shown in lattice
units as a function of L (horizontal coordinate). 
The insert on the left shows the avoided level crossings at sharp 
resonance locations
with the choice ${\rm \lambda =10}$. On the right, with ${\rm \lambda =0}$, 
all the resonances melted away into broad scattering states.
}
\label{fig:fig4}
\vskip -0.3in
\end{figure}
%
%
%
Consider a massive scalar field
in one space and one time dimension interacting with an external source
J(x) according to the Lagrangian 
${\rm L = \frac{1}{2}\partial_\mu\Phi\partial_\mu\Phi
- \frac{1}{2}M^2\Phi^2 - J(x)\Phi^2 }$. The field is
confined between two opaque barriers 
represented by repulsive delta-function potentials
${\rm J(x) = -\frac{1}{2}M^2(\Theta(x) - \Theta(x-L))+}$ 
${\rm \lambda(\delta (x)+\delta (x-L)) }$ 
where $\lambda$ is tunable and the field is rendered massless inside
a square well.
The field eigenmodes ${\rm E_n(L) }$ and ${\rm C_{eff}(L)}$ can be 
calculated as shown
in Fig.~\ref{fig:fig4}. There are only four
bound "string modes" for the two choices of $\lambda$.
On the left side ${\rm \lambda=10}$, ${\rm C_{eff}=0.99}$ and the sharp 
resonance spectrum is located at the expected string positions. On the right,
${\rm \lambda=0}$ and the sharp resonances disappear from 
the bulk scattering state
spectrum. Nevertheless, ${\rm C_{eff}=0.85}$ is only 15 percent off from the
${\rm L=\infty}$ limit. A hidden mechanism on phase shifts might keep 
the conformal charge close to what it would be for a perfect
massless string spectrum trapped inside.


This work was supported by the U.S.~DOE, Grant No. DE-FG03-97ER40546, 
the U.S. National Science Foundation under Award PHY-0099450 and the European
Community's Human Potential Programme, HPRN-CT-2000-00145.

\end{document}